\documentclass[aps,twocolumn,superscriptaddress,prb,amsmath]{revtex4-2}
\usepackage{multirow}
\usepackage[latin9]{inputenc}
\usepackage{mathrsfs}
\usepackage{txfonts}
\usepackage{amssymb}
\usepackage{graphicx,subfigure,float}
\usepackage{dcolumn}
\usepackage{bbm}
\usepackage{bm}
\usepackage{color}
\usepackage[colorlinks, linkcolor=blue, citecolor=blue, urlcolor=blue]{hyperref}
\usepackage{lipsum}
\usepackage{makecell}
\usepackage{gensymb}
\usepackage{pifont}
\usepackage{placeins}
\begin{document}

\newcommand{\Ez}{E_z}
\newcommand{\Da}{D_a}
\newcommand{\dk}{\mathrm{d}^2 k}

\title{Prediction of a layer nonlinear Hall effect in bilayer nonmagnetic or antiferromagnetic systems}

\author{Zehou Li}
\affiliation{Key Laboratory of Micro-nano Energy Materials and Application Technologies, University of Hunan Province $\&$ College of Physics and Electronic Engineering, Hengyang Normal University, Hengyang 421002, China}
\author{Shenda He}
\affiliation{School of Science, Hunan Institute of Technology, Hengyang 421002, China}
\author{Pan Zhou}
\email{zhoupan71234@xtu.edu.cn}
\affiliation{Key Laboratory of Low Dimensional Materials and Application Technology of Ministry of Education,
School of Materials Science and Engineering, Xiangtan University, Xiangtan 411105, China}
\author{Baoru Pan}
\affiliation{Hunan Provincial Key laboratory of Thin Film Materials and Devices, School of Materials Science
and Engineering, Xiangtan University, Xiangtan 411105, China}
\author{Rui Tan}
\email{rtan@hynu.edu.cn}
\affiliation{Key Laboratory of Micro-nano Energy Materials and Application Technologies, University of Hunan Province $\&$ College of Physics and Electronic Engineering, Hengyang Normal University, Hengyang 421002, China}
\author{Lizhong Sun}
\email{lzsun@xtu.edu.cn}
\affiliation{Key Laboratory of Low Dimensional Materials and Application Technology of Ministry of Education, School of Materials Science and Engineering, Xiangtan University, Xiangtan 411105, China}

\date{\today}

\begin{abstract}
Nonlinear Hall effects provide a powerful probe of quantum geometry in solids and enable rectification phenomena beyond the constraints of linear response. In this Letter, we predict a \emph{layer nonlinear Hall effect} (LNHE) in stacked bilayer systems composed of nonmagnetic or antiferromagnetic materials with a vanishing linear Hall conductivity. In such systems, the second- or third-order nonlinear Hall responses are intrinsically layer odd: the contributions from the two constituent layers have equal magnitude but opposite sign, resulting in exact cancellation under layer-exchange symmetry. An out-of-plane electric field $E_z$ can break this symmetry, thereby unveiling the hidden response and converting it into a switchable macroscopic nonlinear Hall signal. Using a minimal $k\!\cdot\!p$ model, we demonstrate that the LNHE can originate from the Berry curvature dipole mechanism. A systematic symmetry analysis of all 80 layer groups further yields a complete classification of the symmetry constraints and stacking configurations that allow for this type of LNHE. Beyond this mechanism, additional symmetry analysis reveals that LNHE may also arise from quantum metric dipole or inversed mass dipole. Remarkably, even in cases where second-order nonlinear Hall responses are symmetry forbidden, a third-order LNHE can still survive in certain stacked bilayer configurations. First-principles calculations on representative bilayers---nonmagnetic 1T$'$-WTe$_2$ and 1T$'$-ReS$_2$---explicitly demonstrate electrically reversible second-order LNHE, in full agreement with our symmetry-based predictions. Overall, our results establish LNHE as a universal phenomenon in a wide range of layered quantum materials and provide a robust route toward electrically tunable nonlinear transport.
\end{abstract}

\maketitle

\textit{Introduction}---The layer degree of freedom has recently emerged as an important internal quantum number in layered materials, enabling transport phenomena that go beyond conventional charge- and spin-based responses \cite{layer1,layer2,layer3,layer4,layer5,work1,work2,layer6,layer7,layer8,layer9,layer10,layer11,layer12,layer13,layer14,layer15,layer16,layer17,layer18}. A paradigmatic example is the layer Hall effect, which occurs in bilayer or multilayer systems when Berry curvature becomes selectively localized to individual layers \cite{LHE1,LHE2,LHE3,LHE4,LHE5,LHE6,LHE7,LHE8,LHE9,LHE10,LHE11,LHE12,LHE13,LHE14,LHE15,LHE16,LHE17,LHE18,LHE19}. Consequently, charge carriers localized in different layers acquire opposite anomalous velocities under an in-plane electric field, generating transverse Hall currents of equal magnitude but opposite sign. Although these contributions cancel in the total charge Hall response due to layer-exchange symmetry, the resulting layer-resolved Hall currents persist. These hidden Hall currents encode rich Berry-curvature physics and provide a powerful route for controlling electronic transport through stacking configurations, interlayer coupling, or external electric fields. As such, the layer Hall effect offers a natural platform for probing layer-resolved geometric effects and developing layertronic functionalities.\\
\indent In systems preserving time-reversal symmetry, the conventional linear Hall effect is forbidden, making the second-order nonlinear Hall effect the lowest-order transverse Hall response permitted by symmetry  \cite{NLHE1,NLHE4,NLHE6,NLHE17,NLHE20,BCD1,BCD2,BCD3,BCD4,BCD5,QMD2,NLHE23,QMD1,BCP1,QMD3,QMD5,NLHE10,Drude1,Drude2,Drude3,Drude4,NLHE3,NLHE8,NLHE15,NLHE24,NLHE27,NLHE22,NLHE25,NLHE26,NLHE7,NLHE11,NLVHE,NLHE16,NLHE5,NLHE14,NLHE13,NLHE2}. Microscopically, the nonlinear Hall effect originates from intrinsic band-geometry-related mechanisms, including the Berry curvature dipole (BCD) \cite{NLHE1,NLHE4,NLHE6,NLHE17,NLHE20,BCD1,BCD2,BCD3,BCD4,BCD5}, quantum metric dipole(QMD) \cite{QMD2,QMD3,NLHE23,QMD1,BCP1,QMD5,NLHE10}, as well as inversed mass dipole (IMD) arising from semiclassical dynamics and disorder effects \cite{Drude1,Drude2,Drude3,Drude4}. Experimentally, the nonlinear Hall effect has been observed in a variety of two-dimensional (2D) systems, including transition metal dichalcogenides and antiferromagnetic heterostructure, and is highly sensitive to symmetry, band structure, and external perturbations \cite{QMD4,NLHE2,NLHE5,NLHE13,NLHE14,experiment1,experiment2,experiment3,experiment7,experiment8,experiment9}. Despite these significant advances, most existing studies have focused on monolayer and bulk systems. By contrast, the influence of internal degrees of freedom--most notably the layer index in bilayer structures--remains largely unexplored. This naturally raises the question of whether a layer-dependent counterpart of the nonlinear Hall effect can exist.\\
\indent In this work, we address this question by introducing the concept of the \emph{layer nonlinear Hall effect} (LNHE), which integrates the layer degree of freedom with nonlinear Hall responses in nonmagnetic and antiferromagnetic bilayer systems. We show that, for specific interlayer stacking configurations, both second- and third-order nonlinear Hall responses can be layer-odd: the nonlinear Hall currents generated in the two constituent layers are equal in magnitude but opposite in sign, leading to an exact cancellation enforced by layer-exchange symmetry. An out-of-plane electric field breaks this symmetry, thereby converting the hidden layer-resolved response into a switchable macroscopic nonlinear Hall signal. Using minimal $k{\cdot}p$ and tight-binding models, we explicitly demonstrate the emergence of LNHE driven by the BCD. A systematic symmetry analysis of all 80 layer groups further establishes the general symmetry criteria and stacking configurations that enable LNHE within this mechanism. Beyond the BCD scenario, additional symmetry considerations reveal that LNHE can also originate from QMD or IMD. Notably, we predict a third-order LNHE that survives even when all second-order nonlinear Hall responses are symmetry forbidden. Finally, first-principles calculations for representative nonmagnetic bilayers, 1T$^\prime$-WTe$_2$ and 1T$^\prime$-ReS$_2$, explicitly confirm the existence of LNHE.

\textit{BCD and Layer-Locked Nonlinear Hall Response}---In time-reversal invariant but inversion-broken systems, the Berry curvature $\Omega(\mathbf{k})$ is odd in momentum space and therefore integrates to zero over the Brillouin zone. However, its asymmetric distribution in momentum space allows its first moment of BCD to remain finite. This BCD serves as the geometric origin of the second-order nonlinear Hall response and is defined as \cite{NLHE1}
\begin{equation}
D_{ab} = \sum_n \int \frac{d^d k}{(2\pi)^d}
\, f_n(\mathbf{k}) \, \partial_{k_a} \Omega_{n,b}(\mathbf{k}),
\end{equation}
where $n$ labels the bands, $f_n(\mathbf{k})$ is the Fermi distribution function, and $\Omega_{n,b}(\mathbf{k})$ is the Berry curvature component. For a 2D system, only the out-of-plane Berry curvature $\Omega_z$ is nonzero, and the relevant in-plane BCD components are $D_{yxx}^{BCD}$ and $D_{xyy}^{BCD}$\cite{NLHE1,NLHE17,BCD1}. In the semiclassical regime, a nonzero BCD generates a transverse second-order current under an in-plane electric field, constituting the intrinsic nonlinear Hall effect\cite{NLHE1,NLHE2,BCD1}. Experimentally, this effect is commonly detected through the second-harmonic ($2\omega$) Hall voltage in response to an AC driving field with frequency $\omega$, and the sign of the measured signal directly reflects the sign of the corresponding BCD component \cite{NLHE18,NLHE19,NLHE2,NLHE5,NLHE13,NLHE14,experiment1,experiment2,experiment3,experiment4,experiment5,experiment6,experiment7,experiment8,experiment9,experiment10}.

\begin{figure}[t]
  \centering
  \includegraphics[width=\columnwidth]{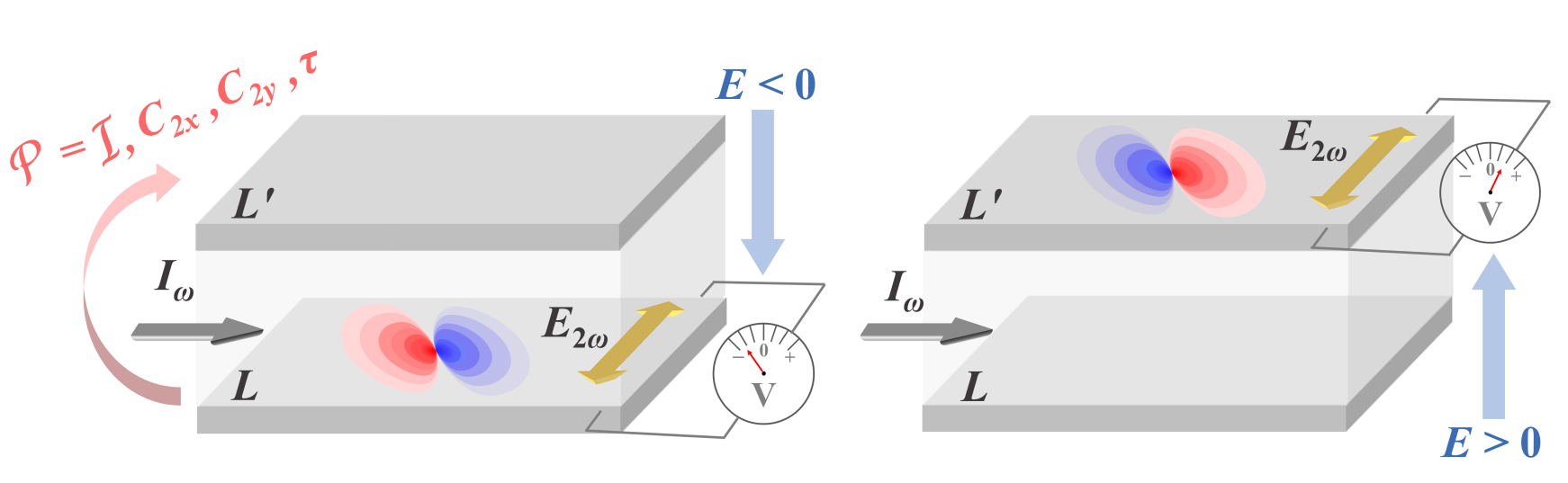}
  \caption{\label{fig:schematic}
Schematic setup and concept of LNHE in a bilayer. An AC in-plane drive at frequency $\omega$ generates a second-harmonic transverse response at $2\omega$, whose polarity is controlled by the out-of-plane bias field $E_z$ (switching between $E_z>0$ and $E_z<0$). In the layer-symmetric configuration the layer-resolved nonlinear Hall signals are opposite and can cancel, while $E_z$ breaks layer equivalence and yields a finite net $2\omega$ signal.}
\end{figure}

In this paper, we interpret the nonlinear Hall response within a layertronic framework. For a constituent monolayer to support a finite second-order Hall current, the point group symmetry must permit a non-vanishing BCD. We extend this to an inversion-connected bilayer system, denoted as $B = L + \mathcal{P}L(L')$, where $\mathcal{P}$ represents the inversion operation mapping layer $L$ to its partner $L'$. From a global symmetry perspective, the bilayer assembly is centrosymmetric, which dictates a vanishing macroscopic BCD tensor. However, this global nullification is a consequence of a layer-locked mechanism: the local nonlinear Hall currents generated within layer $L$ are perfectly compensated by equal and opposite currents in layer $L'$. This represents a hidden degree of freedom where the geometric response is present but latent due to the layer-exchange symmetry. To activate this hidden response, we introduce an external out-of-plane electric field ($E_{\perp}$) as a symmetry-breaking perturbation. Theoretically, $E_{\perp}$ couples to the layer index, establishing a potential gradient that lifts the layer-exchange degeneracy. By tuning the field strength, the low-energy electronic bands of a specific layer can be selectively shifted toward the Fermi level. This lifting of symmetry disrupts the exact compensation between layers $L$ and $L'$, effectively ``unmasking'' the local BCD, as shown in Fig. \ref{fig:schematic}. Consequently, the previously latent layer-locked currents are converted into a measurable macroscopic nonlinear Hall signal. The polarity of the resulting Hall voltage is determined by which layer's contribution dominates the Fermi surface; thus, reversing the electric field direction induces a sign reversal in the transport signal. This model establishes a mechanism for the electrical activation and layer-selective control of nonlinear quantum transport in van der Waals bilayers.

\textit{Minimal bilayer effective model}---For a minimal description of the LNHE, we consider a bilayer system whose low-energy physics is captured by a Hamiltonian defined in the combined space of layer ($\tau_i$) and orbital ($\sigma_i$) degrees of freedom,
\begin{equation}
H(\mathbf{k}) = H_0(\mathbf{k})\tau_z + t_\perp\tau_x + \Delta\tau_z,
\end{equation}
The first term describes two decoupled layers with opposite signs, the second term $t_\perp$ accounts for interlayer hybridization, and $\Delta$ represents the interlayer potential difference induced by an out-of-plane electric field. The single-layer building block $H_0(\mathbf{k})$ is chosen as a minimal two-band model that preserves time-reversal symmetry but breaks inversion symmetry,
\begin{equation}
H_0(\mathbf{k}) = d_x(\mathbf{k}) \sigma_x + d_y(\mathbf{k}) \sigma_y + d_z(\mathbf{k}) \sigma_z,
\end{equation}
where
\[
d_x(\mathbf{k}) = k_x,\quad
d_y(\mathbf{k}) = k_y, \quad
d_z(\mathbf{k}) = m + \alpha k_x .
\]
Here $m$ is a mass term, while the linear term $\alpha k_x$ explicitly breaks inversion symmetry. This inversion breaking skews the Berry-curvature distribution in momentum space and gives rise to a finite BCD in an isolated layer. Finally, the interlayer potential difference is given by $\Delta$ = e$d$$E_z$, with $d$ the interlayer spacing and $E_z$ the applied out-of-plane electric field. The bilayer construction encodes opposite geometric responses on the two layers, yielding a hidden layer-odd nonlinear Hall effect that cancels at $\Delta = 0$ due to layer-exchange symmetry. A finite $\Delta$ redistributes the low-energy spectral weight between layers near the Fermi level, unbalancing the opposite layer BCDs and generating a net BCD, and hence a finite LNHE. Reversing $E_z$ changes the sign of $\Delta$ and thus reverses the polarity of the nonlinear Hall response.

\begin{figure}[t]
  \centering
  \includegraphics[width=\columnwidth]{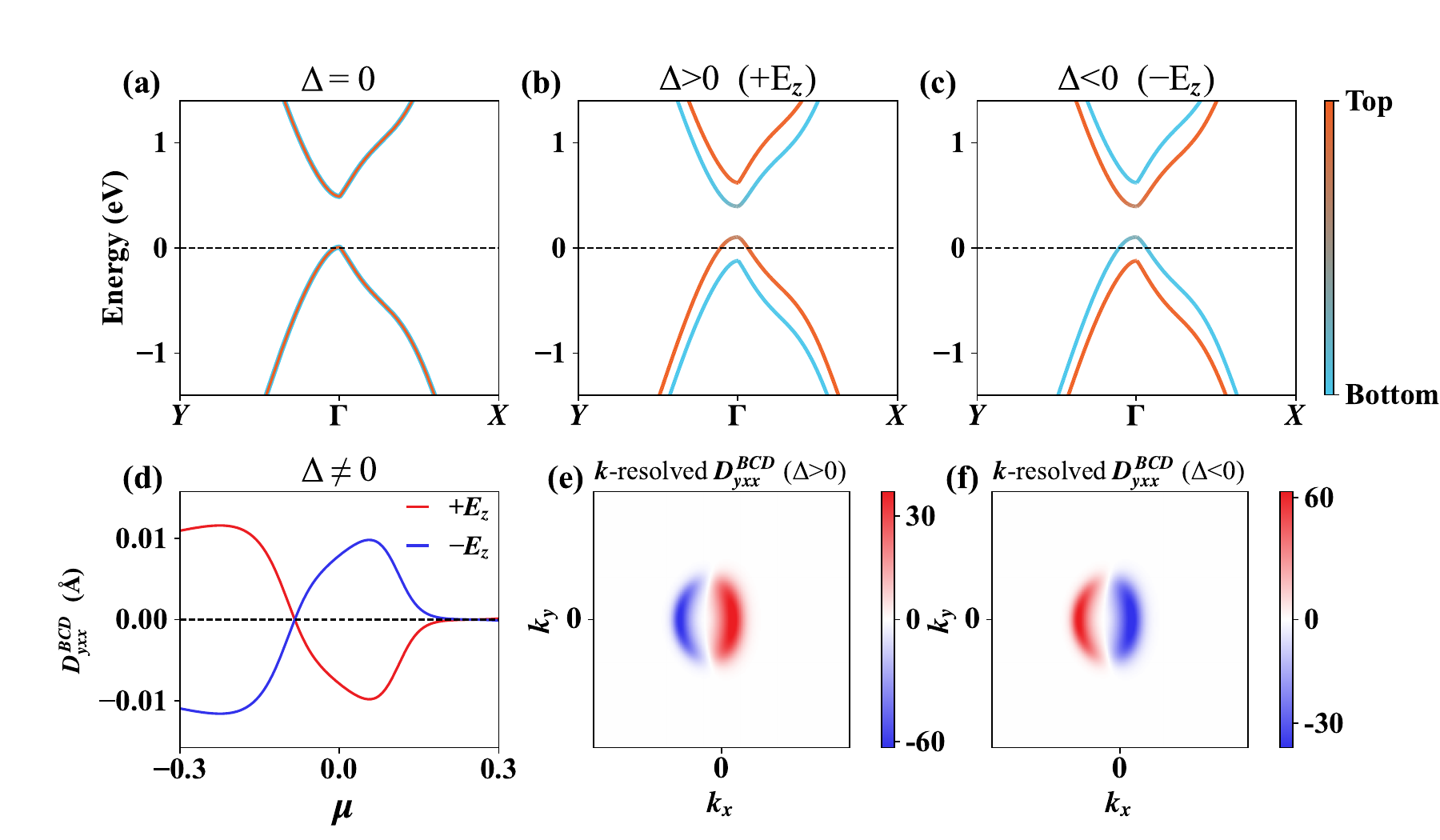}
  \caption{\label{fig:fig2}
Tight-binding results for LNHE. (a)--(c) Band structures for $\Delta = 0$, $\Delta > 0$ ($+E_z$), and $\Delta < 0$ ($-E_z$). Color contrast indicates layer-associated polarization, illustrating how $E_z$ redistributes low-energy spectral weight between the two layers. (d) BCD component $D_{yxx}$ as a function of chemical potential $\mu$ for opposite field directions, showing a clear sign reversal under $E_z \to -E_z$. (e),(f) Momentum-resolved contributions to $D_{yxx}^{BCD}$ for $\Delta > 0$ and $\Delta < 0$ at $\mu$=0.07, demonstrating that the geometric hot spots invert upon field reversal. The parameters used are $\alpha = 0.71, B=0.86, m=0.2, \delta = 0.15, t_\perp = 0.14$.}
\end{figure}

To bridge the continuum physics and the lattice regularization, we map the low-energy $k\cdot p$ theory onto a square-lattice tight-binding Hamiltonian:
\begin{equation}
H_{\mathrm{TB}}(\mathbf{k}) = \sin k_x \,\sigma_x + \sin k_y \,\sigma_y + M(\mathbf{k})\,\sigma_z ,
\end{equation}
where $M(\mathbf{k}) = m + \alpha \sin k_x - 2B(2 - \cos k_x - \cos k_y)$. In the long-wavelength limit ($\mathbf{k} \to \mathbf{0}$), this Hamiltonian recovers the effective continuum model. Based on this tight-binding model, we compute the nonlinear response, and the results are presented in Fig.~\ref{fig:fig2}. At the layer-symmetric point [$\Delta = 0$, Fig.~\ref{fig:fig2}(a)], the global inversion symmetry dictates an exact mutual cancellation of the BCD between the layers, despite the existence of finite local geometric activity. Breaking this layer-exchange symmetry via a finite $\Delta$ [Fig.~\ref{fig:fig2}(b) and \ref{fig:fig2}(c)] unmasks a non-vanishing $D_{yxx}^{BCD}$. As illustrated in Fig.~\ref{fig:fig2}(d), $D_{yxx}^{BCD}(\mu)$ is strictly antisymmetric with respect to the field direction, $E_z \to -E_z$, providing a mechanism for an electrically switchable second-order nonlinear Hall response. This macroscopic switching is microscopically rooted in the field-induced inversion of singular geometric textures---or ``hot spots"---highly localized in reciprocal space [Fig.~\ref{fig:fig2}(e) and \ref{fig:fig2}(f)].

\textit{Symmetry analysis}---Previous studies have established that the second-order nonlinear Hall effect requires extremely low crystalline symmetry, particularly for the BCD-induced response\cite{NLHE1}. In such systems, the crystal may possess at most a single mirror plane or a single $C_2$ rotation; the presence of any additional symmetry of this type forces the nonlinear Hall conductivity to vanish. Motivated by this constraint, we formulate the symmetry analysis directly for the field-polarized configuration with $E_z\neq 0$, which explicitly breaks all bilayer symmetries that reverse $z$ to $-z$. Under these conditions, a wide class of systems can, in principle, support a nonzero second-order nonlinear Hall conductivity.

\indent The layer-switchable nonlinear Hall effect imposes two essential requirements. First, the field-polarized bilayer must support a finite in-plane BCD, which is the prerequisite for a nonlinear Hall response. This condition constrains the symmetries that remain after applying the out-of-plane electric field: the symmetry of each monolayer may contain at most a single vertical mirror plane or a single in-plane $C_2$ rotation, or neither of the two. Any additional symmetry would prohibit the emergence of an in-plane BCD under $E_z$. Second, reversing the external electric field, $E_z \to -E_z$, must reverse the sign of the relevant in-plane BCD component, thereby enabling electrical switching of the nonlinear Hall signal. For 2D transport, symmetry allows only two independent in-plane BCD components, namely $D_{yxx}^{BCD}$ and $D_{xyy}^{BCD}$. This requirement reflects a geometric constraint intrinsic to the bilayer structure. Specifically, the $+E_z$ and $-E_z$ configurations must be related by a symmetry operation that simultaneously exchanges the two field polarities via $z \to -z$ and reverses the sign of the in-plane nonlinear Hall response, i.e., $D_{yxx}^{BCD}$ and/or $D_{xyy}^{BCD} \to -(D_{yxx}^{BCD}, D_{xyy}^{BCD})$. A systematic point-group analysis for 2D crystals shows that only three symmetry operations satisfy both criteria: $C_2^{x}$, $C_2^{y}$, and spatial inversion $\mathcal{I}$. Consequently, LNHE switching can be realized by engineering a stacking configuration whose point-group component contains one of these operations, thereby guaranteeing a BCD sign reversal upon electric-field reversal.

\indent The perpendicular electric field $\Ez$ plays two roles. It breaks the layer-exchange symmetry, thereby allowing the hidden layer-locked nonlinear Hall response to become observable. At the same time, $\Ez$ can also modify the band structure, Berry curvature, and BCD. We analyze the total BCD calculated from the finite-$\Ez$ Hamiltonian. In the weak-field regime, the total BCD follows the symmetry-allowed expansion $\Da(\Ez)=\alpha_a \Ez+\gamma_a\Ez^3+O(\Ez^5)$. If the band structure, Fermi surface, and Berry-curvature distribution change only weakly with $\Ez$, the electric field mainly acts to break layer-exchange symmetry and unmask the hidden layer-locked response. A strong nonlinear field dependence or substantial reconstruction of the Berry-curvature distribution would instead indicate that the field directly modulates the BCD.

\indent These symmetry constraints naturally lead to two generic stacking strategies. \emph{Class I} employs an inversion-related stacking of the two monolayers, which guarantees the required sign reversal under $E_z \rightarrow -E_z$. Within this class, an in-plane relative translation between the monolayers is allowed. Such a translation can preserve the inversion correspondence of the bilayer under $E_z \rightarrow -E_z$, while breaking other crystalline symmetries that are intrinsically robust against an external electric field and therefore cannot be lifted by $E_z$ alone. This flexibility is particularly important for stacked bilayers: after applying a vertical electric field, any residual symmetry operations beyond mirror or $C_2$ symmetries can be readily eliminated by introducing an interlayer translation. \emph{Class II} preserves a single vertical mirror symmetry $\sigma_\perp$ under $E_z$ by choosing the stacking translation within the mirror-invariant plane, while the sign reversal is enforced by $C_2^{x}$ or $C_2^{y}$. Since $\sigma_\perp$ neither inverts $z$ nor exchanges inequivalent layers at different heights, \emph{Class II} additionally requires the monolayer itself to possess the same $\sigma_\perp$ symmetry. Based on the general stacking theory previously proposed in Ref.~\cite{work1}, we provide a complete enumeration of all 80 layer groups and their admissible stacking registries, including the allowed monolayer symmetries and stacking prescriptions, in Tab.~S1 of the Supplemental Material (SM)\cite{SM}.

\begin{figure}[t]
  \centering
  \includegraphics[width=\columnwidth]{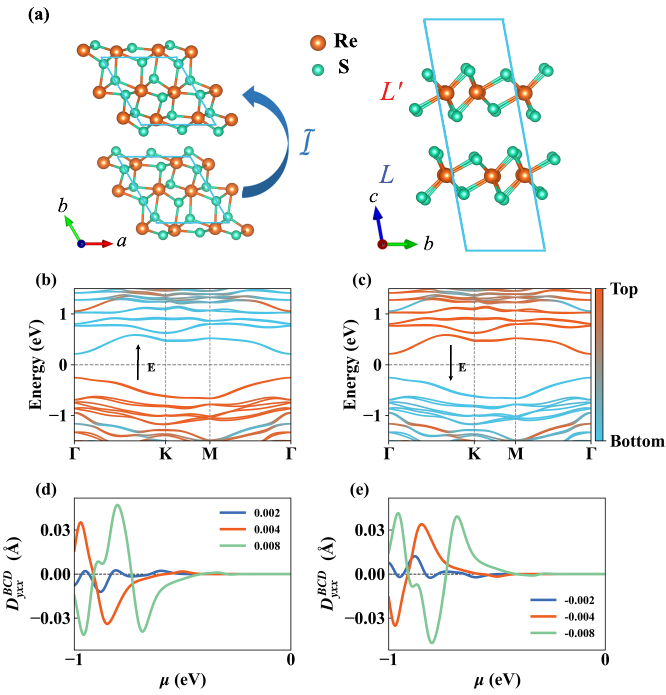}
  \caption{\label{fig:fig3}
Material realization of LNHE. (a) Bilayer 1T$^\prime$-ReS$_2$ with an inversion-related stacking configuration. (b),(c) Layer-resolved band structures along high-symmetry lines for opposite out-of-plane electric fields, showing a field-induced layer polarization of the low-energy states. (d),(e) BCD component $D_{yxx}^{BCD}$ as a function of chemical potential for opposite field directions, exhibiting  a clear sign reversal under $E_z \to -E_z$.}
\end{figure}

\begin{table}[t]
\centering
\small
\setlength{\tabcolsep}{4pt}
\renewcommand{\arraystretch}{1.25}
\caption{\label{tab:magnetic_summary}
Summary of in-plane nonlinear Hall channels in magnetic systems.
Column 2 shows the subset that can remain compatible with a finite out-of-plane electric field $E_z$.
Column 3 lists stacking operations (the point-group part of the bilayer stacking operator) that can enforce a sign reversal of the given in-plane channel between $+E_z$ and $-E_z$ by mapping $z\!\to\!-z$ and flipping the corresponding response component.}
\begin{tabular}{lcc}
\hline\hline
\makecell[l]{In-plane \\ component} &
\makecell[c]{Operations} &
\makecell[c]{Stacking \\ operation} \\
\hline
$D^{\mathrm{BCD}}_{yxx}$ & $\{\sigma_x, \mathcal{T}, \sigma_x\mathcal{T}\}$ & $\{C_2^{x}\}$ or $\{\mathcal{I}\}$ \\[2pt]
$D^{\mathrm{BCD}}_{xyy}$ & $\{\sigma_y, \mathcal{T}, \sigma_y\mathcal{T}\}$ & $\{C_2^{y}\}$ or $\{\mathcal{I}\}$ \\[4pt]

$D^{\mathrm{QMD}}_{yxx}$ & $\{\sigma_x, C_2^z\mathcal{T}, \sigma_y\mathcal{T}\}$ & $\{C_2^{x}\}$ or $\{\mathcal{I}\}$ \\[2pt]
$D^{\mathrm{QMD}}_{xyy}$ & $\{\sigma_y, C_2^z\mathcal{T}, \sigma_x\mathcal{T}\}$ & $\{C_2^{y}\}$ or $\{\mathcal{I}\}$ \\[4pt]

$D^{\mathrm{IMD}}_{yxx}$ & $\{\sigma_x, C_3^z, S_6^z, C_2^z\mathcal{T}, C_6^z\mathcal{T}, \sigma_y\mathcal{T}, S_3^z\mathcal{T}\}$ & $\{C_2^{x}\}$ or $\{\mathcal{I}\}$ \\[2pt]
$D^{\mathrm{IMD}}_{xyy}$ & $\{\sigma_y, C_3^z, S_6^z, C_2^z\mathcal{T}, C_6^z\mathcal{T}, \sigma_x\mathcal{T}, S_3^z\mathcal{T}\}$ & $\{C_2^{y}\}$ or $\{\mathcal{I}\}$ \\
\hline\hline
\end{tabular}
\end{table}

\indent In addition to the BCD-induced second-order nonlinear Hall effect discussed above, two other representative mechanisms--the QMD and the IMD--can also give rise to nonlinear Hall responses. Based on their respective symmetry requirements, we systematically analyze the layer-resolved nonlinear Hall effect arising from these two mechanisms. All symmetry-allowed bilayer operations and stacking configurations are summarized in Tab.~\ref{tab:magnetic_summary}. These results demonstrate that, for suitable monolayer symmetries and stacking prescriptions, the layer-resolved nonlinear Hall effect can be realized via both QMD and IMD mechanisms (see details in \cite{SM}). In addition to non-magnetic 2D materials, certain antiferromagnetic 2D monolayers also exhibit a lowest-order response that is second order\cite{NLHE10,NLHE16,NLHE23}. Unlike non-magnetic systems, whose symmetry properties are fully described by ordinary crystallographic point groups, antiferromagnetic materials are characterized by magnetic point or space groups, in which the relevant symmetry operations are combinations of crystal symmetries and time-reversal symmetry. Consequently, these kinds of operation are also included in our symmetry analysis. In the nonmagnetic bilayer systems considered here, time-reversal symmetry $\mathcal{T}$ is preserved. Since the BCD contribution is $\mathcal{T}$-even whereas the QMD and IMD contributions are $\mathcal{T}$-odd (see Tab.~S3 in the SM\cite{SM}), the LNHE in the present systems originates solely from the Berry curvature dipole. For magnetic systems where $\mathcal{T}$ is broken, all three mechanisms may coexist and can be distinguished by their characteristic relaxation-time scalings, $\chi_{\mathrm{BCD}}\propto \tau$, $\chi_{\mathrm{QMD}}\propto \tau^0$, and $\chi_{\mathrm{IMD}}\propto \tau^2$, as illustrated in Sec.~II of the Supplemental Material.

\indent Beyond the second-order response, recent studies have shown that symmetry constraints in many 2D materials suppress second-order contributions, making the third-order Hall response the leading term and giving rise to a third-order nonlinear Hall effect \cite{third1,third2,third3,third4,third5,third6,NLHE12,NLHE9}. Our layer-resolved nonlinear Hall effect framework can be naturally extended to this third-order regime. To explicitly illustrate this extension, we construct a bilayer altermagnetic model, in which the Hall conductivity exhibits opposite responses under opposite vertical electric fields, and is dominated by the electronic states of a single layer. Detailed discussions are provided in Sec.~III of the SM\cite{SM}.

\textit{Material realization}---Guided by the symmetry analysis above, we propose that an experimentally synthesized 1T$^\prime$-WTe$_2$ bilayer \cite{NLHE5,NLHE14} with $C_2^x$ stacking can host a layer-dependent BCD-induced nonlinear Hall effect, whose sign can be reversed by switching the direction of the vertical electric field (see Sec. IV of the SM \cite{SM} for details). In light of recent advances in the controlled fabrication of bilayer 2D materials with tailored stacking configurations, experimental observation of this effect should be within reach ; a detailed discussion of experimental realization steps is given in the SM\cite{SM}. Although 1T$^\prime$-WTe$_2$ constitutes a realistic \emph{Class II} platform, the absence of a sizable band gap near the Fermi level implies that, under an applied out-of-plane electric field, the Fermi surfaces at different chemical potentials generally receive contributions from both layers. This hybridization complicates an unambiguous layer-resolved identification of the nonlinear Hall conductivity.

\indent To achieve a clearer layer-resolved identification and simultaneously illustrate the complementary \emph{Class I} mechanism, we next consider experimentally synthesized 1T$^\prime$-ReS$_2$ \cite{ReS21,ReS22,ReS23}. Monolayer 1T$^\prime$-ReS$_2$ belongs to the layer group $\mathrm{p}\bar{1}$ (No.~2). We construct the bilayer by inversion-related stacking $\mathcal{I}$, as shown in Fig.~\ref{fig:fig3}(a).  This configuration renders the bilayer centrosymmetric with point group $C_i$, which enforces a vanishing macroscopic BCD. Upon applying an out-of-plane electric field, inversion symmetry $\mathcal{I}$ is broken and the symmetry is reduced to $C_1$, thereby generically allowing a finite BCD. Importantly, within a broad energy window around the Fermi level, the electronic states [Figs.~\ref{fig:fig3}(b) and \ref{fig:fig3}(c)] are predominantly localized on one of the two layers, evidencing strong electric-field-induced layer polarization. Correspondingly, the nonlinear Hall conductivity [Figs.~\ref{fig:fig3}(d) and \ref{fig:fig3}(e)] exhibits opposite signs at selected Fermi levels when the direction of the applied electric field is reversed. Combined with the projected band structures, this sign reversal provides unambiguous evidence that the nonlinear Hall response mainly originates from a single layer, thereby realizing the LNHE. Note that, we calculated interlayer sliding energy barriers and phonon dispersions for bilayer 1T$'$-ReS$_2$, which confirm its dynamical stability (see Sec.~VI of the the SM\cite{SM} for details). Furthermore, within the considered chemical potential range, the maximal magnitude of the nonlinear Hall conductivity increases monotonically with electric field strength, consistent with the enhanced layer polarization and the concomitant growth of the BCD under stronger inversion-symmetry breaking.

\textit{Discussion}---In summary, by analyzing the symmetry properties of bilayer systems, we conceptually establish the LNHE as an intrinsic consequence of bilayer symmetry, thereby linking the spatially discrete layer degree of freedom to momentum-space quantum geometric characteristics. Based on a systematic group-theoretical analysis, we further identify the symmetry constraints required for the emergence of the LNHE under different physical mechanisms. These results highlight the LNHE as a powerful probe of layer polarization and a versatile means for electrically controlling nonlinear transport. In light of the experimental realization of monolayer 1T$^\prime$-ReS$_2$ and 1T$^\prime$-WTe$_2$, together with rapid advances in bilayer stacking techniques, we expect our theoretical predictions to be experimentally accessible in the near future.

\indent Although the sign reversal of the nonlinear Hall signal under an out of plane electric field is macroscopically consistent with symmetry in any inversion symmetric system, the layer degree of freedom brings essential microscopic advantages that go far beyond a generic realization of that principle. 
First, the layer index acts as a discrete bipartite quantum degree of freedom--a ``layer pseudospin''--to which $E_z$ couples like a Zeeman field. 
This coupling drives a discrete quantum transition from layer hybridized to layer polarized states, in contrast to the continuous electron cloud distortion (the Stark effect) that occurs in ordinary bulk materials. 
Second, individual monolayers possess low symmetry geometric properties (the Berry curvature dipole, quantum metric dipole, or inverse mass dipole) that are perfectly compensated in a centrosymmetric bilayer and therefore remain intrinsically hidden. 
By selectively addressing the layer pseudospin, $E_z$ activates the internal quantum geometry of a chosen atomic layer, and reversing the field transfers the active conduction channel entirely to the opposite layer. 
This selective activation of hidden geometric dipoles, and the resulting spatial transfer of the transport channel, are hallmarks of layertronics that have no counterpart in conventional bulk systems. 
Third, while previous layer Hall studies were largely confined to the linear response of magnetic materials, our framework generalizes layertronics to the nonlinear regime and unifies it across both magnetic and nonmagnetic van der Waals bilayers. 
By demonstrating that higher order geometric tensors can be electrically controlled and layer switched in a vast family of materials, we establish nonlinear layertronics as a comprehensive paradigm for future quantum device design. A systematic comparison of the LNHE with the layer Hall effect is presented in Sec.~VII of the SM\cite{SM}.

\indent From a device perspective, the LNHE offers a promising platform for layer-resolved electronic functionalities in low-dimensional systems. Owing to its direct sensitivity to layer polarization and symmetry-controlled nonlinear response, the LNHE can be exploited to design electrically tunable rectifiers, nonlinear Hall detectors, and layer-selective signal processing elements without the need for magnetic fields. Moreover, the strong dependence of the LNHE on stacking configuration and external electric fields enables flexible device engineering through structural design and gating.

\indent Combined with the availability of high-quality van der Waals bilayers and mature fabrication techniques, these features position LNHE-based devices as attractive candidates for next-generation low-power nonlinear electronics.

We note that a recent related preprint \cite{arXiv23971} discusses an out-of-plane-field-tunable layer-resolved nonlinear Hall effect in $\mathcal{PT}$-symmetric antiferromagnetic insulators. A detailed comparison with that work is provided in Sec.~VIII of the SM\cite{SM}.

\begin{acknowledgments}
\indent This work is supported by the National Natural Science Foundation of China (Grant No. 12574070 and Grant No. 12504223), the China Postdoctoral Science Foundation(2025M773383 and GZC20252231), the China Postdoctoral Science Foundation-Hunan Joint Support Program(2025T002HN), the Scientific Research Fund of Hunan Provincial Education Department (24B0658), Science Foundation of Hengyang Normal University of China (Grant No. 2024QD38).\\
\end{acknowledgments}

\nocite{abacus1,abacus2,GGA,ONCV,SG15,NAO,PYATB,wan2022,liu2025,fei2018,ma2019,cucchi2018}
\bibliographystyle{apsrev4-2}
\indent Zehou Li, Shenda He and Pan Zhou contributed equally to this work.
\bibliography{references}

\end{document}